\documentclass[aps,epsfig,showpacs,amssymb,floatfix]{revtex4}
\usepackage{bm} \usepackage{graphicx}
\usepackage{citesort}

\newcommand{\be}{\begin{equation}}
\newcommand{\en}{\end{equation}}
\newcommand{\bear}{\begin{eqnarray}}
\newcommand{\enar}{\end{eqnarray}}

\newcommand{\paa}{\partial}

\newcommand{\veps}{\varepsilon}

\newcommand{\dd}{\mbox{d}}

\newcommand{\la}{\langle}
\newcommand{\ra}{\rangle}

\newcommand{\br}{{\bf r}}

\newcommand{\bu}{{\bf u}}

\newif\ifpdf\ifx\pdfoutput\undefined\pdffalse\else\pdfoutput=1\pdftrue\fi

\begin{document}
\title{Lagrangian statistics of particle pairs in homogeneous
isotropic turbulence}

\author{L. Biferale$^{1}$, G. Boffetta$^{2}$, A. Celani$^{3}$,
B.J. Devenish$^{1}$, A. Lanotte$^{4,*}$, and F. Toschi$^{5}$}

\affiliation{$^1$ Dipartimento di Fisica, Universit\`a ``Tor Vergata",
and INFN, Sezione di Roma II, Via della Ricerca Scientifica 1,
I-00133 Roma, Italy} 
\affiliation{$^2$ Dipartimento di Fisica
Generale and INFN, Universit\`a di Torino, Via P.Giuria 1, I-10125
Torino, Italy}
\affiliation{$^3$ CNRS, INLN, 1361 Route des Lucioles, F-06560 Valbonne, France} 
\affiliation{$^4$ CNR, ISAC, Str. Prov. Lecce-Monteroni km. 1200 , I-73100 Lecce, Italy}
\affiliation{$^5$ CNR, IAC, Viale del Policlinico 137, I-00161 Roma, Italy}
\affiliation{$^*$ Corresponding author\,: a.lanotte@isac.cnr.it}

\begin{abstract}
We present a detailed investigation of the particle pair separation
process in homogeneous isotropic turbulence. We use data from direct
numerical simulations up to $R_{\lambda} \sim 280$ following the
evolution of about two million passive tracers advected by the flow
over a time span of about three decades. We present data for both the
separation distance and the relative velocity statistics. Statistics
are measured along the particle pair trajectories both as a function
of time and as a function of their separation, i.e. at fixed
scales. We compare and contrast both sets of statistics in order to
gain an insight into the mechanisms governing the separation
process. We find very high levels of intermittency in the early
stages, that is, for travel times up to order ten Kolmogorov time
scales. The fixed scale statistics allow us to quantify anomalous
corrections to Richardson diffusion in the inertial range of scales
for those pairs that separate rapidly. It also allows a quantitative
analysis of intermittency corrections for the relative velocity
statistics.
\end{abstract}
\pacs{47.27.-i  47.10.+g}
\maketitle

\section{Introduction}
The relative dispersion of pairs of particles is important because of
its connection with the problem of concentration fluctuations
\cite{Batchelor,Durbin,Thomson} and because of the insight it provides
into the spatial structure of turbulent flows. In contrast with single
particle dispersion, which is mostly driven by the large scale --
energy containing -- eddies, the dispersion of pairs of particles
depends on velocity fluctuations of order the separation of the pairs.
Thus, the early stages of relative dispersion, up to the integral time
scale, are expected to reflect the universal nature of small scale
turbulence (independent of the large scale flow) and the intermittent
character of the energy cascade. The latter appears to manifest itself in
some particle pairs remaining close together for long periods of time
while others separate rapidly.

Clearly, a good understanding of the mechanisms of relative dispersion
will lead to better models. Among key features of relative dispersion
are a separation dependent time scale and long-time correlations of
quantities such as the relative velocity. Many different types of
quantitative models of relative dispersion have been proposed
including Lagrangian stochastic models
e.g. \cite{Thomson,Kurba,Borgas} and kinematic simulation
\cite{vass1,vass2}. For a review of relative dispersion and Lagrangian
stochastic models in particular, we refer the reader to Sawford
\cite{Sawford}. In recent years it has become apparent that the
success of these models at small scales will depend on their ability
to capture the intermittency of the separation process
\cite{Borgas}. One purpose of this paper is to provide a detailed
quantitative and qualitative analysis of the separation process which
we hope will eventually lead to improved models.

Results from observations of the spread of marked particles (pairs or
clouds of tracers in the atmosphere and in the ocean), summarised in
classical textbooks such as those by Monin and Yaglom \cite{Monin} and
Pasquill and Smith \cite{Pasquill}, is testimony to the difficulty in
getting reliable experimental data in fully developed
turbulence. Although much progress has been made in recent years in
experimental measurements of single Lagrangian particles
\cite{LaPorta,Mordant}, relatively few Lagrangian measurements have
been obtained following pairs of particles. A notable exception is Ott
and Mann \cite{Ott} who report Lagrangian inertial range scaling even
at modest Reynolds numbers, of the order $R_{\lambda} \sim 100$, where
$R_{\lambda}$ is the Taylor scale Reynolds number. As a result, direct
numerical simulation (DNS) is still the most important source of
detailed Lagrangian statistics
(e.g. \cite{Yeung,Yeung2,Boff,Ishihara,Yeung4} at Reynolds numbers up
to order $R_{\lambda}\sim 280$).  In this paper we analyse the results
of a recent DNS of $3-D$ homogeneous isotropic
turbulence seeded with Lagrangian particles \cite{Bif}. Although
homogeneous isotropic turbulence has limited application to real
situations, it is the simplest configuration for studying the
statistics of relative dispersion.

The paper is organised as follows. In section
\ref{DNS} we outline the numerical scheme used for calculating the
data and discuss statistical uncertainty and variability. We present
results on the statistical properties of both the particle pair
separation and its relative velocity. These are considered in sections
\ref{sep_stats} and \ref{vel_stats}, respectively. In both cases we
compute the statistics as a function of time and as a function of
separation, that is, at fixed scales. The latter allows for a more
accurate separation of the dissipative, inertial and the integral
scale regions.
\section{DNS methodology} 
\label{DNS}
The direct numerical simulation of homogeneous isotropic turbulence
was performed on $512^3$ and $1024^3$ cubic lattices with Reynolds
numbers $R_\lambda \sim 180$ and $R_\lambda \sim 280$, respectively.
The Navier-Stokes equations were integrated using fully de-aliased
pseudo-spectral methods for a total time spanning nearly three decades
(from the order of a tenth of the Kolmogorov time scale, $\tau_\eta$,
to approximately three times the integral time scale, $T_L$). The flow
was forced by keeping the total energy constant in the first two
wavenumber shells \cite{Chen}. The flow at $R_\lambda=284$ was seeded
with approximately two million Lagrangian passive tracers once a
statistically stationary velocity field had been obtained. The
positions and velocities of the particles were stored at a sampling
rate of $0.07 \tau_\eta$. The numerical parameters are summarised in
Table~\ref{tab1}. In this DNS, dissipative scales are well resolved,
satisfying $\eta \sim \dd x$, where $\dd x$ is the grid spacing. The
Lagrangian velocity was calculated using linear interpolation which
was demonstrated to be adequate for obtaining well resolved particle
accelerations \cite{Bif2}.

The particles were initially arranged in tetrads which were uniformly 
distributed in the flow. A total of 960,000 pairs with initial 
separations $r_0 = 1.2\eta$ and $r_0 = 2.5\eta$ were formed this way. 
Particle pairs with larger initial separations were formed by 
following pairs chosen from different tetrads. In this way we also 
follow pairs with initial separations $r_0 = 9.8\eta$ and $r_0 = 19.6\eta$.  
The number of pairs varied from $5.10^5$ to $1.10^6$ depending on the 
chosen initial separation. In both cases a particle may have been used
more than once to form a pair. 

Previous studies have shown that Lagrangian statistics are affected by
highly non-Gaussian fluctuations (see e.g. \cite{Yeung4,Boff}). Thus,
it is important to quantify the statistical uncertainty of some
typical variable in order to ensure the reliability of the results
within our statistical sample. Statistical errors were estimated by
dividing the sample into five sub-ensembles and calculating the
minimum and maximum values. We find that the error is at worst
approximately $15\%$ for the separation skewness and approximately
$25\%$ for the relative velocity skewness. The nature of the forcing
scheme used in the present DNS meant that relatively little temporal
variability of globally-averaged quantities was observed (see Overholt
and Pope \cite{Overholt} for discussion on forcing schemes and
temporal variability). In particular, fluctuations about the mean of
the energy dissipation, $\veps$, were at most 10\% during the
evolution of the DNS. Thus, in the following we may safely use $\veps$
(and other globally-averaged quantities) to scale the two-particle
statistics.
\section{Separation statistics} \label{sep_stats}
\subsection{Fixed time statistics} \label{sep_time}

We consider the motion of two marked fluid particles, labelled by the
superscripts $(1)$ and $(2)$. In homogeneous turbulence, it is
sufficient to consider the statistics of the instantaneous separation
of the positions of the two particles, namely ${\bf r}(t) = {\bf
r}^{(1)}(t) - {\bf r}^{(2)}(t)$. Furthermore, in isotropic turbulence,
the separation magnitude $r=|{\bf r}|$ plays a fundamental role in the
problem of relative dispersion.

Following the well known ideas of Richardson \cite{Richardson}, 
relative dispersion in the inertial range of time scales, 
$\tau_{\eta} \ll t \ll T_{L}$, can be described in
terms of a diffusion equation for the probability density function
(pdf) of the pair separation $p({\bf r},t)$. In spherical coordinates
this is given by
\begin{equation} 
\frac{\paa p({\bf r},t)}{\paa t} = \frac{1}{r^2} \frac{\paa}{\paa r}
\! \left( r^2 K(r) \frac{\paa p({\bf r},t)}{\paa r} \right),
\label{eq:diffusion}
\end{equation} 
where $K(r)$ is a scalar eddy diffusivity. On the basis of
experimental measurements in the atmosphere, Richardson proposed that
$K(r) = k_0 \veps^{1/3} r^{4/3}$ where $k_0$ is a dimensionless
constant. Assuming a small enough initial separation and a large
enough travel time, it can be shown (see e.g. Monin and Yaglom
\cite{Monin}, p. 574) that a spherically symmetric solution of
(\ref{eq:diffusion}) is given by
\begin{equation} 
p(r,t) = \frac{A r^2}{(k_0 \veps^{1/3} t)^{9/2}} \exp\!\left(
-\frac{9 r^{2/3}}{4 k_0 \veps^{1/3} t} \right),
\label{sep_pdf}
\end{equation} 
where $A=(3/2)^8/\Gamma(9/2)$ is a normalisation constant. This
exhibits strong non-Gaussianity with a narrow peak at the origin and
very large tails and gives rise to the celebrated scaling for the
second order moment
\begin{equation} 
\la r^2 \ra = g \veps t^3.
\label{eq:t3}
\end{equation} 
Here $g=1144 k_0^3/81$ is the Richardson constant which is supposed to
be universal. This result was also derived by Obukhov \cite{Obukhov}
using Kolmogorov's classical theory of turbulence (K41) \cite{Monin}.

The Richardson pdf is perfectly self-similar, all positive moments
behave according to the dimensional law: $r^p \propto t^{3p/2}$.
The scaling (\ref{eq:t3}) is notoriously difficult to achieve both in
laboratory experiments and in DNS on account of the large separation
of scales that is required to observe it. As a result, estimates of
$g$ have varied widely, from $0.06$ to $3.5$ \cite{Sawford}. The main
practical difficulties in achieving a long inertial subrange are due
to dissipative range effects at the ultraviolet end of the spectrum
and integral scale effects at the infrared end of the spectrum. In
the dissipation range, pairs separate exponentially and with widely
varying growth rates -- some pairs separate rapidly while others
remain close together. This leads to the formation of a broad
distribution of separations. As a result, slowly separating pairs
(which remain in the dissipative range) and rapidly separating pairs
(which approach the integral scales) \lq contaminate' the statistics
in the inertial range.  A very large Reynolds number is therefore
required to produce reliable Lagrangian statistics in the inertial
range.

\begin{figure}
\includegraphics[scale=0.6]{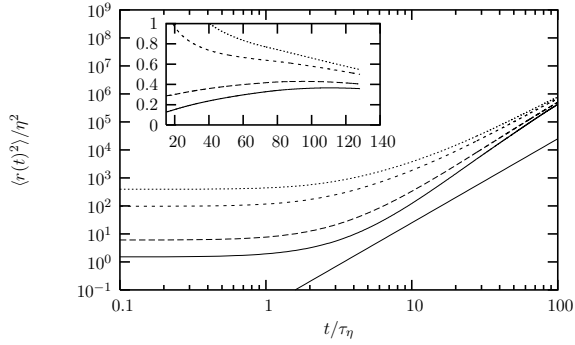}
\caption{
  The evolution of $\la r(t)^2 \ra/ \eta^2 $ vs $t/\tau_{\eta}$ for
  the initial separations $r_0=1.2\eta$, $r_0=2.5\eta$, $r_0=9.8
  \eta$, and $r_0=19.6\eta$. The straight line is proportional to
  $t^3$.  Inset: $\la r(t)^2 \ra/\veps t^3$ for the same four initial
  separations starting from $t/\tau_{\eta} \sim 15$.  }
\label{fig:r2_mean_1}
\end{figure}

In Figure \ref{fig:r2_mean_1} we plot the mean square separation $\la
r^2 \ra$ versus $t$, normalised by the Kolmogorov microscales, $\eta$
and $\tau_{\eta}$ respectively. Although the curves begin to collapse
at large $t$, they do not display a $t^3$ scaling and still show a
dependence on the initial separation. Thus, any attempt to extract the
value of the Richardson constant will be marred by the memory of the
initial separation. The simplest way to measure $g$ is to plot $\la
r^2 \ra$ scaled by the asymptotic prediction, $\veps t^3$, and look
for a plateau. These curves are displayed in the inset of Figure
\ref{fig:r2_mean_1}. It is clear that none of them produces a good
plateau and given the spread of curves with different initial
separations, the value will be at best an order of magnitude estimate
subject to considerable uncertainty.

An alternative method, used in \cite{Yeung4,Ishihara,Ott}, consists of
fitting a straight line to $\la r^2 \ra^{1/3}$ in a suitable time
interval. If equation (\ref{eq:t3}) holds, this straight line, when
extrapolated back towards $t=0$, should pass through the origin and
have a slope of $(g \veps )^{1/3}$. For all curves we find a small
non-zero intercept whose value varies with $r_0$. This introduces an
extra free parameter in the linear fit corresponding to the non-zero
intercept. The curve with the smallest non-zero intercept has
$r_0=2.5\eta$ and gives a value of $g=0.47$ with an error of the order
of approximately $10\%$ depending on the time range (here taken to be
$15 \tau_{\eta} \le t \le 75 \tau_{\eta}$).  This value of $g$ is
smaller than that found by Yeung and Borgas \cite{Yeung4} and
Ishihara and Kaneda \cite{Ishihara}, though still of the same order of
magnitude, but agrees well with that of Ott and Mann \cite{Ott} and
Boffetta and Sokolov \cite{Boff}.

In order to make a more complete analysis of Richardson's model, we
compute the pdf of the separation distance. The Richardson pdf relies
on two phenomenological assumptions: the first is that the
eddy-diffusivity is self-similar, the second is that the velocity
field is short-time correlated.  However, it is known that anomalous
corrections to the K41 scalings exist (see e.g. \cite{Frisch}) and
these are likely to complicate the situation.

\begin{figure}
\includegraphics[scale=0.6]{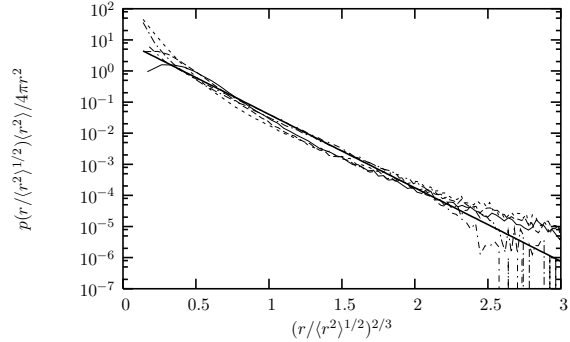}
\caption{
  Comparison of the Richardson pdf with the DNS data. The curves refer
  to data for $r_0=1.2\eta$ at $t=5.2 \tau_\eta$ (solid line), $t=7
  \tau_\eta$ (long dashed line), $t=14 \tau_\eta$ (short dashed line),
  $t=42 \tau_\eta$ (dotted line) and $t=70 \tau_\eta$ (dot-dashed
  line). The thick solid line is the Richardson pdf (\ref{sep_pdf}).
}
\label{fig:pdf_richardson}
\end{figure}

In Fig.~\ref{fig:pdf_richardson} we compare the separation pdf for the
smallest initial separation, $r_0=1.2\eta$, calculated from the DNS
data, with that predicted by Richardson, namely (\ref{sep_pdf}). For
small times (up to $t \sim 40 \tau_{\eta}$), we observed a rapid
change in shape with the pdf showing a pronounced tail, which
indicates that while most pairs are still close together some have
moved very far apart (not shown). At these times the curves do not
rescale indicating that the early stages of the separation process is
very intermittent. Here, the physics of the dissipative range still
exerts an influence on the separation process and so we would not
expect agreement with the Richardson pdf. Only for times in the range
$40-70\tau_\eta$ do we find reasonable agreement with the Richardson
pdf. We note that while at $t \sim 40 \tau_{\eta}$ we find good
agreement for the tail but a large mismatch for values close to the
peak, at $t \sim 70\tau_{\eta}$ the pdf is almost undistinguishable
from (\ref{sep_pdf}). At large times the particles are moving more or
less independently and so the pdf of $r^2$ will be a chi squared
distribution with three degrees of freedom (not shown).

\begin{figure}
\includegraphics[scale=0.6]{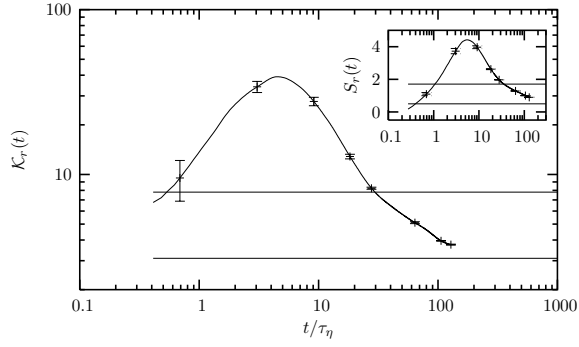}
\caption{
  The separation kurtosis for the smallest initial separation
  $r_0=1.2\eta$. Also shown at some times are the error bars
  calculated from the minimum and maximum values of the five
  sub-ensembles.  Inset: the separation skewness for the same initial
  separation together with the error bars at the same times. The
  horizontal lines are the appropriate values derived from the
  Richardson pdf and the chi squared distribution with three degrees
  of freedom. These values are $1.7$ and $0.49$, respectively, for the
  skewness and $7.81$ and $3.1$, respectively, for the kurtosis.  }
\label{fig:skewness_kurtosis_r}
\end{figure}

A more detailed analysis of the separation pdf can be made by
considering the separation skewness, $S_r(t) = \langle (r(t) -
\overline{r(t)})^3\rangle/(\sigma^2_r(t))^{3/2}$, and the kurtosis,
${\cal K}_r(t) = \langle (r(t) - \overline{r(t)})^4
\rangle/(\sigma^2_r(t))^2$, where $\overline{r}$ is the mean
separation distance and $\sigma_r$ is the root mean square separation
distance. These are shown in Fig.~\ref{fig:skewness_kurtosis_r} for
$r_0=1.2\eta$ and clearly show the intermittent nature of the
separation process at small times. The Richardson pdf, of course,
predicts constant values for the skewness and kurtosis, namely $1.7$
and $7.81$ respectively and which are not reached until approximately
$t \sim 35\tau_\eta$.  This time is within the inertial subrange and
we may have expected the skewness and kurtosis to level off before
decreasing to their large time values ($0.49$ and $3.1$ respectively).
That this is clearly not the case suggests that either contamination
of the inertial range due to the dissipative and integral scales
prevents us from having a region of constant skewness and kurtosis, or
points to shortcomings in the Richardson model.

These results put the difficulties of calculating Richardson's
constant (described above) into context. A perfect collapse of curves
in the pdf would have implied self-similarity but its absence is not
necessarily an indication of the failure of the Richardson model; as
we have already discussed, each end of the inertial range is affected
by, respectively, dissipation range and integral scale effects.  In
section \ref{exit-times} we show how these problems may be overcome by
measuring statistics at fixed scales.

\begin{figure}
\includegraphics[scale=0.6]{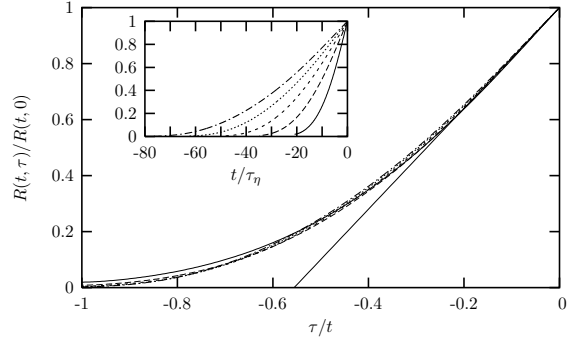}
\caption{
  The nor\-ma\-li\-zed cor\-re\-la\-tion fun\-cti\-on
  $R(t,\tau)/R(t,0)$ versus $\tau/t$ for $r_0=1.2\eta$ at different
  travel times. Inset: the same correlation functions now plotted
  versus $t/\tau_{\eta}$. Curves (from left to right) refer to travel
  times $t=77\tau_\eta$, $t=63 \tau_\eta$, $t=49 \tau_\eta$,
  $t=35\tau_\eta$ and $t=21 \tau_\eta$.  }
\label{fig:r_correlation}
\end{figure}

We conclude this section by considering the correlation function
$R(t,\tau)=\la r(t)r(t+\tau)\ra$ of the separation distance for travel
times within the inertial subrange. This quantity, which probes two
different times along the separation process, is influenced by the
temporal properties of the turbulent energy cascade sustaining the
separation growth. In the inset of Fig.~\ref{fig:r_correlation} we
plot $R(t,\tau)$ for $-t\le \tau \le 0$ at different travel times $t$
for pairs with initial separation $r_0=1.2\eta$. In agreement with
Jullien {\em et al.} \cite{Jullien}, we find that $R(t,\tau)$ broadens
with increasing travel time indicating that the pairs decorrelate more
slowly at larger travel times, a consequence of the fact that larger
and larger eddies have slower and slower dynamics.  In the body of the
figure we plot the same data versus $\tau/t$.  Dimensional analysis
shows that $R(t,\tau)/R(t,0) = f(\tau/t)$.  We see that all the curves
collapse onto a single curve supporting the dimensional prediction.
\subsection{Fixed scale statistics}
\label{exit-times}
To disentangle the effects of different scales, an alternative
approach, based on {\it exit time} statistics, has been proposed
\cite{ABCCV}. This consists of fixing a set of thresholds, $r_n =
\rho^n r_0$, where $\rho>1$ and $n=1,2,3,\ldots$ and then calculating
the time, $T$, taken for the pair separation to change from $r_n$ to
$r_{n+1}$. By averaging over the particle pairs, we obtain the mean
exit time, $\la T_{\rho}(r_{n}) \ra$, or mean {\it doubling time} if
$\rho=2$. Formally we are calculating the first passage time. The
advantage of this approach is that all pairs are sampled at the same
scales and that finite Reynolds number effects are less important
\cite{ABCCV}. For particle pairs with initial condition $p({\bf
r},0)=\rho^2 \delta(r-r_n/\rho)/4\pi r_n^2$, a perfectly reflecting
boundary condition at $r=0$, and an absorbing boundary condition at
$r=r_n$, the pdf of the exit time, $T$, is defined to be
\begin{equation}
{\cal P}_{\rho,r_n}(T) =
 -\frac{\dd}{\dd T} \int_{|{\bf r}|<r_n} p({\bf r},T) \, \dd {\bf r}, \nonumber
\end{equation}
from which we get
\begin{equation}
{\cal P}_{\rho,r_n}(T) = -4 \pi k_0\veps^{1/3} r_n^{10/3}
\left. \frac{\paa p}{\paa r} \right|_{r=r_n},
\label{eq:pdf_exit}
\end{equation} 
on making use of (\ref{eq:diffusion}). Following Boffetta and Sokolov
\cite{Boff2}, we can derive a solution of the $3D$ diffusion equation
(\ref{eq:diffusion}) in terms of an eigenfunction decomposition. This
gives us
\begin{equation}
p(\xi,t) = \sum_{i=1}^{\infty} c_i \exp(-\lambda_i^2 t) \xi^{-7/2} J_{7/2}(3
\lambda_i \xi), \nonumber
\end{equation}
where $\xi=(k_0 \veps^{1/3})^{-1/2} r^{1/3}$, $J_{7/2}(x)$ is a
Bessel function of the first kind, $\lambda_i = 1/3 \, (k_0
\veps^{1/3})^{1/2} r^{-1/3} j_{7/2,i}$ where $j_{7/2,i}$ are the zeros
of $J_{7/2}(x)$ and $c_i$ are constants.  It then
follows from (\ref{eq:pdf_exit}) that the large time asymptotic form
of the exit time pdf is given by
\begin{equation}
{\cal P}_{\rho,r_n}(T) \sim \exp \! \left( -\kappa \, \frac{\rho^{2/3} -
1}{\rho^{2/3}} \frac{T}{\la T_{\rho}(r_n) \ra} \right),
\label{eq:pdf_exit_asym}
\end{equation}
where $\kappa\approx 2.72$ is a numerical constant derived from the
leading zero of the Bessel function described above.

\begin{figure}
\includegraphics[scale=0.6]{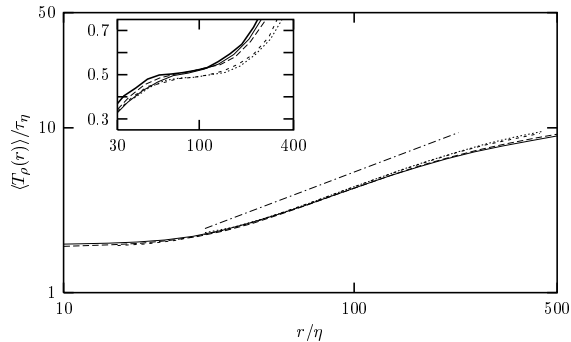}
\caption{
  The mean exit time for the initial separations $r_0=1.2\eta$ (thin
  continuous line), $r_0=2.5\eta$ (long-dashed line), $r_0=9.8 \eta$
  (short-dashed line) and $r_0=19.6\eta$ (dotted line) with
  $\rho=1.25$. The straight line is proportional to $r^{2/3}$. In the
  inset we show Richardson's constant, $g$, versus $r/\eta$ as given
  by (\ref{richardson_const}) for the same initial separations at
  $R_{\lambda}=284$. To evaluate the variability of $g$ with the
  Reynolds number, we also plot a curve (thick continuous line) for
  the initial separation $r_0=1.2\eta$ at $R_{\lambda} = 183$.  }
\label{fig:mean_exit_time}
\end{figure}

Using Richardson's diffusion equation (\ref{eq:diffusion}), the mean exit
time can be shown to be \cite{Boff}
\begin{equation}
\la T_{\rho}(r_n) \ra = \frac{1}{2 k_0} \frac{\rho^{2/3} -
1}{\rho^{2/3}} \frac{r_n^{2/3}}{\veps^{1/3}} \,.
\label{eq:mean_double_time}
\end{equation} 
In the body of Fig.~\ref{fig:mean_exit_time} we plot $\la
T_{\rho}(r_n) \ra$ for a range of initial separations.  It is
immediately clear that there is no dependence on the initial
separation in contrast to the mean square separation calculated as a
function of time (see Fig.~ \ref{fig:r2_mean_1}). Moreover, we see a
clear inertial scaling region in which the mean exit time grows almost
like $r^{2/3}$.

Equation (\ref{eq:mean_double_time}) provides us with a method for
calculating the Richardson constant (since $k_0$ is related to $g$):
\begin{equation} 
g = \frac{143}{81} \frac{ (\rho^{2/3} - 1)^3 }{\rho^2}
\frac{r^2}{\veps \la T_\rho(r) \ra^3}\,.
\label{richardson_const}
\end{equation} 
In the inset of Figure \ref{fig:mean_exit_time}, we plot the
expression (\ref{richardson_const}) for the Richardson constant versus
$r$, for various initial conditions. We see that a collapse of curves
is beginning to form for all initial separations.  We estimate the
value of $g$ to be approximately $0.50 \pm 0.05 $, which agrees with
the value computed above and with previous estimates of $g$
\cite{Ott,Boff,Gioia}. This method has the advantage of relative
insensitivity to the initial separation and avoids the problem of the
non-zero intercept discussed in \ref{sep_time}.  Of course, the
present calculation of $g$ assumes the validity of Richardson's model.
We find that $g$ does not change significantly for $\rho \in
[1.15,2]$. It is also worth noticing that the estimate of $g$ is not
very sensitive to the Reynolds number (see inset of
Fig.~\ref{fig:mean_exit_time}).

\begin{figure}
\includegraphics[scale=0.6]{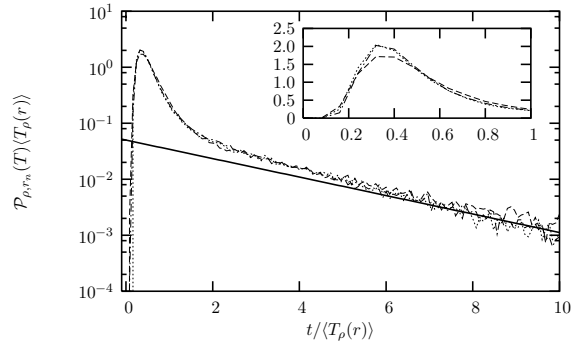}
\caption{
  The log-linear plot of the exit time pdf for $r_0=1.2\eta$ with
  $\rho=1.25$ at $r=21.8 \eta$ (dashed line), $r=83.3\eta$ (dotted
  line), and $r=130.1\eta$ (dot-dashed line). The solid line is the
  large time prediction (\ref{eq:pdf_exit_asym}). Inset: a lin-lin
  plot of the same figure showing more detail.  }
\label{fig:pdf_exit}
\end{figure}

The exit time pdf, ${\cal P}_{\rho,r_n}(T)$, is shown in
Fig.~\ref{fig:pdf_exit} for $r_0=1.2\eta$ and clearly shows the
exponential nature of the exit time pdf at large times. Moreover, it
shows a clear collapse of curves for intermediate and large exit
times, $T_{\rho}(r_n) \ge \langle T_{\rho}(r_n) \rangle$, indicating
that the exit time statistics in this range are self-similar. We note
here that this collapse deteriorates with increasing $\rho$. We have
shown that by focusing on properties at fixed scales a good agreement
with the theoretical prediction (\ref{eq:pdf_exit}) can be
achieved. Free of the effects of the infrared and ultraviolet
cut-offs, the Richardson diffusion model appears to work well for the
inertial range of scales. For small exit times, $T_{\rho}(r_n) \ll
\langle T_{\rho}(r_n) \rangle$, on the other hand, we do not find a
complete collapse of curves at different thresholds, indicating that
pairs with a fast separation are likely to exhibit intermittency (see
inset of Fig.~\ref{fig:pdf_exit}).

\begin{figure}
\includegraphics[scale=0.6]{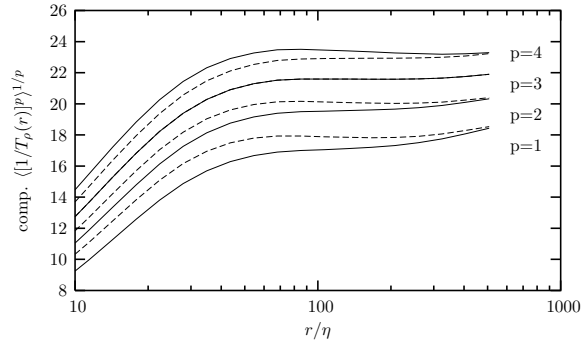}
\caption{
  The inverse exit time moments, $\la [1/T_\rho(r)]^p \ra^{1/p}$, for
  $p=1,\ldots,4$ compensated with the Kolmogorov scalings (solid
  lines) and the multifractal predictions (dashed lines) for the
  initial separation $r_0 = 1.2\eta$ and for $\rho=1.25$.  }
\label{fig:exit_moments}
\end{figure}

The higher order moments of $T$ are dominated by those pairs which
separate slowly. Conversely, the moments of the inverse exit times,
$\la [1/T_{\rho}(r)]^p \ra$, are dominated by those pairs which
separate rapidly and correspond to positive moments of the
separation. Kolmogorov scaling based on dimensional analysis then
leads to
\begin{equation}
\left \la \left( \frac{1}{T_{\rho}(r)} \right)^p \right \ra \sim
\veps^{p/3} r^{-2p/3}.
\label{k41}
\end{equation}
Assuming that a reasonable estimate of the exit time is given by 
$T(r) \sim r/u_r$, where $u_r$ is the relative velocity at scale $r$, 
intermittency corrections can be quantified in terms of the 
multifractal formalism \cite{BCCV}:
\begin{equation}
\left \la \left( \frac{1}{T_{\rho}(r)} \right)^p \right \ra \sim
\frac{1}{T_L^p} \left( \frac{r}{L_0} \right)^{\zeta_E(p) - p},
\label{inte} 
\end{equation}
where $\zeta_E(p)$ are the scaling exponents of the Eulerian velocity
structure functions as predicted by the multifractal formalism. In
Fig.~\ref{fig:exit_moments} we plot $\la \left( 1/T_{\rho}(r)
\right)^p \ra^{1/p}$ scaled by the Kolmogorov (\ref{k41}) and
intermittent scaling exponents (\ref{inte}), respectively. The
$\zeta_E(p)$ are calculated using the She-L\'ev\^eque formula
\cite{She}. As already remarked at lower Reynolds numbers by Boffetta
and Sokolov \cite{Boff}, there is a small but clear improvement in the
scaling of the inverse exit times when scaled by the multifractal
predictions.

\begin{figure}
\includegraphics[scale=0.6]{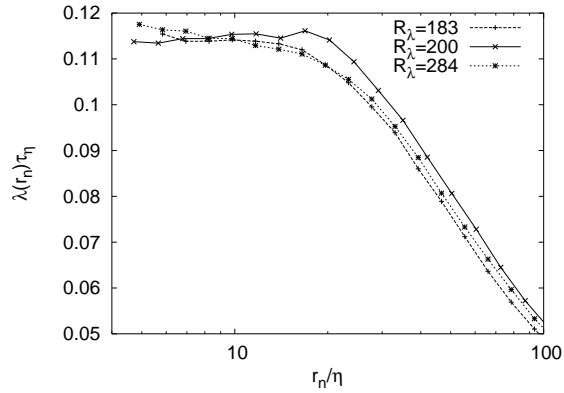}
\caption{
  The finite size Lyapunov exponents as a function of the separation
  $r_n$ for different Reynolds numbers.  }
\label{fig:lyapunov}
\end{figure}

Before concluding this section we note that the exit time statistics
can be used to measure the largest Lyapunov exponent in the flow.
This is because for small thresholds, $r_n$, the mean exit time probes
the exponential growth of the separation distances. The exact relation
between the \lq finite size Lyapunov exponent' and the mean exit time
is \cite{Aurell}
\begin{equation}
\lambda = \lim_{r_n \rightarrow 0} \frac{1}{\langle T_{\rho}(r_n)\rangle}
\log(\rho).
\label{eq:lyapunov}
\end{equation}
In Fig.~\ref{fig:lyapunov} we show the right hand side of
(\ref{eq:lyapunov}) for three different Reynolds numbers (two from
this numerical simulation, see Table 1) and one from a previous DNS
study \cite{Boff}, at different thresholds, $r_n$. The usual Lyapunov
exponent is recovered from the saturation value in the limit of small
$r_n$.  As may be seen in the figure, the data show a clear
proportionality between the Kolmogorov time, $\tau_{\eta}$, and the
Lyapunov exponents, $\lambda$, for all available Reynolds
numbers. Thus, we get
\[
\lambda \tau_{\eta} \sim 0.115 \pm 0.005.
\]
This value is comparable with the one found by Girimaji and Pope
\cite{Girimaji}.

\section{Relative velocity statistics} \label{vel_stats}
\subsection{Fixed time statistics}

We now consider the statistics of the relative velocity of the
particle pairs during the separation process and which we denote as
$\bu_r(t)=\bu^{(1)}(t) - \bu^{(2)}(t)$.  The relative velocity
statistics are of interest because they provide information on the
rate of separation of the particle pairs.  We consider the statistics
of the relative velocity projected in the direction of the separation
vector, the \lq longitudinal' component, and the projection of the
relative velocity orthogonal to the separation, the \lq transverse'
component. The former is given by
\[
u_{||} = \frac{\dd r}{\dd t} = {\bf u}_r \cdot \hat{\bf r},
\]
where $\hat{\bf r} = {\bf r}/r$.  The transverse component of the relative
velocity is given by
\[
{\bf u}_{\bot} = {\bf u}_r - u_{||} \hat{\bf r}.
\]
There are, of course, two transverse components of the relative
velocity but since the turbulence is isotropic it suffices to consider
only one. We comment here that the relative magnitudes of $\la |\bu_r|
\ra$, $\la u_{||} \ra$ and $\la |{\bf u}_{\bot}| \ra$ and the
alignment properties of $\bu_r$, $\br(t)$ and $\br(0)$ have been
discussed extensively by Yeung and Borgas \cite{Yeung4}. Here, we
state simply that our data give similar results and concentrate on
the pdfs of the velocity components and their properties.

\begin{figure}
\includegraphics[scale=0.6]{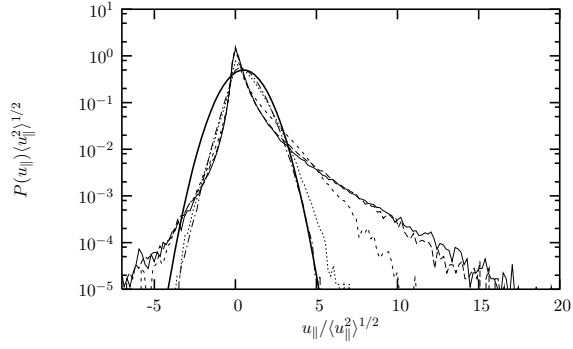}
\caption{The pdf of $u_{||}$ for $r_0=1.2\eta$ at the following
  travel times: (from outer to inner curve) $t=5.2 \tau_\eta$, $t=7
  \tau_\eta$, $t=14 \tau_\eta$, $t=42 \tau_\eta$ and $t=70 \tau_\eta$.
  The thick solid line is a Gaussian distribution.  }
\label{fig:pdf_u_para}
\end{figure}

\begin{figure}
\includegraphics[scale=0.6]{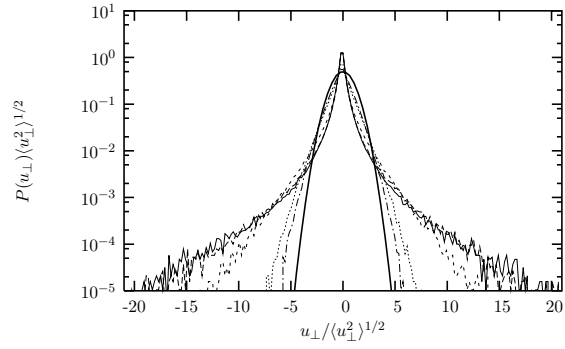}
\caption{The pdf of $u_{\bot z}$ for $r_0=1.2\eta$ at the same times
  as Fig.~\ref{fig:pdf_u_para}. The thick solid line is a Gaussian
  distribution.  }
\label{fig:pdf_u_perp}
\end{figure}

In Fig.~\ref{fig:pdf_u_para} we plot the pdf of the longitudinal
component of the relative velocity, $u_{||}(t)$, for $r_0=1.2\eta$.
The pdf is negatively skewed at $t=0$ (not shown), corresponding to
the Eulerian distribution, but as $t$ increases, it quickly becomes
positively skewed indicating that pairs with small initial separation
are more likely to be diverging than converging.  This skewness then
decreases and the pdf tends towards a Gaussian distribution for travel
times of order $T_L$. The pdf of one component of ${\bf u}_{\bot}$ for
the same initial separation is shown in
Fig.~\ref{fig:pdf_u_perp}. Unlike the pdf of $u_{||}$, it is symmetric
about the mean. Thus, negative velocities are as common as positive
velocities indicating that there is no preferred direction of rotation
as may be expected in isotropic and parity invariant turbulence. We
note here that for both longitudinal and transverse pdfs we do not see
a complete collapse of curves for times in the range $t \in [10, 70]
\tau_{\eta}$.

\begin{figure}
\includegraphics[scale=0.6]{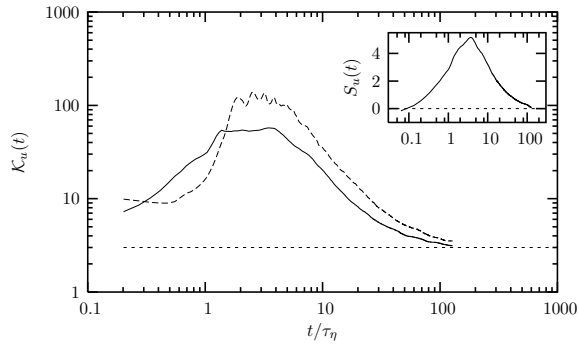}
\caption{The kurtosis of $u_{||}$ (continuous line) and of $u_{\bot}$
  (dashed line) for the smallest initial separations $r_0=1.2\eta$.
  Inset: the skewness of $u_{||}$ for the same initial separation. The
  horizontal lines are the Gaussian values for the kurtosis and
  skewness.  }
\label{fig:u_skewness_kurtosis}
\end{figure}

\begin{figure}
\includegraphics[scale=0.6]{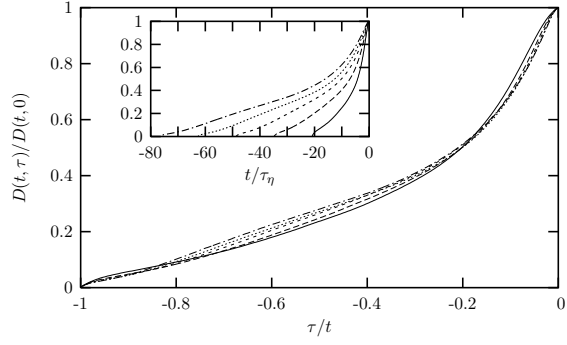}
\caption{The normalised correlation function $D(t,\tau)/D(t,0)$
  versus $\tau/t$ and $t/\tau_\eta$ (inset) for pairs of initial
  separation $r_0=1.2\eta$. Curves (from left to right) refer to
  travel times $t=77\tau_\eta$, $t=63 \tau_\eta$, $t=49 \tau_\eta$,
  $t=35\tau_\eta$ and $t=21 \tau_\eta$.  }
\label{fig:u_correlation}
\end{figure}

We consider the pdfs of the relative velocity components in more
detail by analysing their skewness $S_{u}(t)$ and kurtosis ${\cal
K}_u(t)$. These are shown in Fig.~\ref{fig:u_skewness_kurtosis} for
$r_0=1.2\eta$. At $t=0$ the Lagrangian statistics (not shown) are
identical to the Eulerian statistics. This is reflected in the
negative skewness of $u_{||}$ which is close to $-0.55$, the value
commonly observed for Eulerian velocity structure functions at
moderate to high Reynolds numbers \cite{Sreenivasan}. We also note
that at early times (up to $t \sim 10 \tau$) the maximum values of the
skewness and kurtosis of the velocity statistics are higher than the
corresponding maxima of the separation statistics. We conclude this
section by measuring the correlation of the relative velocity along
the particle pair trajectories. In the inset of
Fig.~\ref{fig:u_correlation} we plot $D(t,\tau)=\la u_{||}(t)
u_{||}(t+\tau) \ra$ for $-t\le \tau \le 0$ for pairs with initial
separation $r_0=1.2\eta$.  In agreement with
Fig.~\ref{fig:r_correlation} we find that $D(t,\tau)$ broadens with
increasing travel time confirming that the velocity decorrelates more
slowly at larger travel times. In the body of the same figure we plot
the same data rescaled versus $\tau/t$. We note here that the
rescaling does not give as good a collapse as for the separation
distance, a fact which suggests the presence of {\it tiny} anomalous
fluctuations in the characteristic times governing the decorrelation
of eddies of different size.
\subsection{Fixed scale statistics}
\label{exit-vel}
Following the exit time method of section \ref{exit-times}, we
calculate the relative velocity at fixed scales in order to achieve 
\lq uncontaminated' inertial range statistics and which we term the 
{\it exit velocities}. We compute the value of the relative 
velocity components
$u_{||}(r)$ and ${\bf u}_{\bot}(r)$ whenever a particle pair has a
separation within a specified logarithmic shell of radius $r = r_n(1
\pm 0.1)$, with $r_n = \rho^n r_0$. This differs from the method we
used to calculate the exit times above as here we are calculating not
just the velocity at the first passage but also at all subsequent
passages.

\begin{figure}
\includegraphics[scale=0.6]{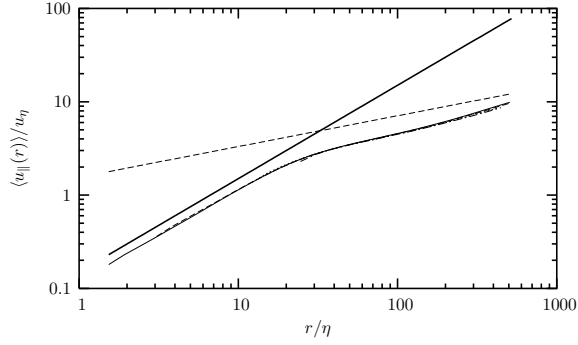}
\caption{The mean longitudinal exit velocity, $\la u_{||} \ra$, as a
  function of the separation and scaled by $u_{\eta`}\eta/\tau_\eta$,
  for pairs of initial separations $r_0=1.2\eta$, $r_0=2.5\eta$,
  $r_0=9.8 \eta$ and $r_0=19.6\eta$ with $\rho=1.25$. The thick solid
  line is proportional to $r$ and the dashed line is proportional to
  $r^{1/3}$.  }
\label{fig:mean_u_r_space}
\end{figure}

In Fig.~\ref{fig:mean_u_r_space} we plot the mean longitudinal exit
velocity, $\la u_{||}(r) \ra$, as a function of the absolute
separation, $r$, for different $r_0$. The lack of dependence on $r_0$
is immediately apparent. We also see a clear separation between
dissipative and inertial range scales, with $\la u_{||} \ra$ growing
linearly for small $r$, corresponding to the dissipation range and
then growing close to $r^{1/3}$ for larger $r$ which corresponds to
the inertial range. Similar behaviour was observed for $|{\bf
u}_{\bot}(r)|$. We note here that the statistics showed little
qualitative change for different values of $\rho$. It is tempting to
think that because of the lack of dependence on the initial separation
and the qualitative agreement with the K41 scalings for Eulerian
statistics, the exit velocities resemble the Eulerian velocity
structure functions even if they are not strictly Eulerian.  However,
the very existence of a non-vanishing mean longitudinal relative
velocity tells us that the two sets of statistics cannot be identical.
We consider these differences in more detail by analysing the pdfs of
the exit velocities.

\begin{figure}
\includegraphics[scale=0.6]{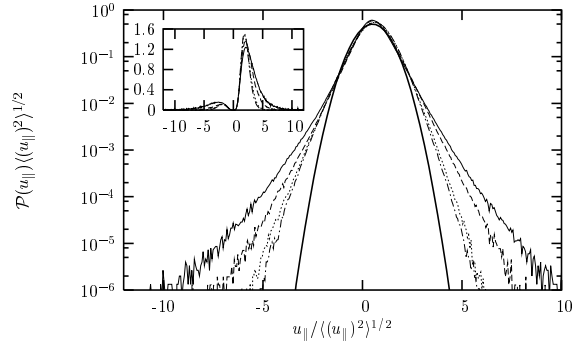}
\caption{The log-lin plot of the exit velocity pdf ${\cal P}(u_{||})$
  calculated as a function of $r$ for pairs with $r_0=1.2\eta$ and
  $\rho=1.25$. The curves are for the following thresholds
  $r=5.72\eta$, $r=21.8\eta$, $r=83.3\eta$ and $130.1\eta$ (from outer
  to inner curve). The thick solid line is a Gaussian distribution.
  Inset: a lin-lin plot of $u_{||}^4 {\cal P}(u_{||})$.  }
\label{fig:pdf_u_r_space}
\end{figure}

\begin{figure}
\includegraphics[scale=0.6]{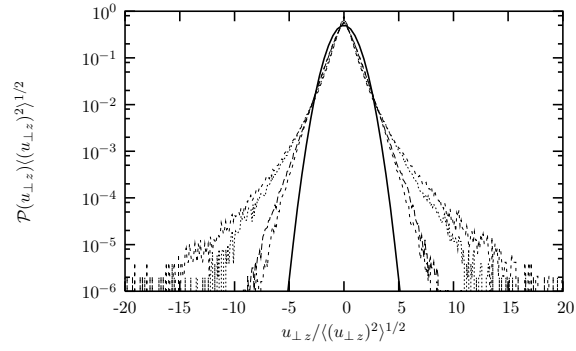}
\caption{The log-lin plot of the exit velocity pdf ${\cal P}(u_{\bot z})$
  for $r_0=1.2\eta$ with $\rho=1.25$, at the same thresholds as
  Fig.~\ref{fig:pdf_u_r_space}. The thick solid line is a Gaussian
  distribution.  }
\label{fig:pdf_u_p_space}
\end{figure}

In Fig.~\ref{fig:pdf_u_r_space} we plot the pdf of the longitudinal
(exit) velocity, ${\cal P}(u_{||})$, and in the inset we plot $u_{||}^4
{\cal P}(u_{||})$. In contrast to the Eulerian pdf, the pdf is slightly
positively skewed initially but as the separation threshold increases,
it becomes more symmetric and tends towards a Gaussian
distribution. However, unlike the pdf of $u_{||}(t)$ (see
Fig.~\ref{fig:pdf_u_para}) we do not see an initial rapid increase in
the positive tail of the pdf. At fixed separations there is not the
spread of contributions to the velocity as there is at fixed
times. The pdf of one component of the transverse relative exit
velocity, ${\cal P}(u_{\bot z})$, is shown in Fig.~\ref{fig:pdf_u_p_space}
for the same initial separation $r_0=1.2\eta$. As may be expected,
this pdf is initially symmetric and remains so for increasing
separation threshold. We note that there is no qualitative change in
the behaviour of both pdfs for different values of $\rho$. The absence
of a complete collapse of curves at different thresholds in both pdfs
indicates that the exit velocities are intermittent.

\begin{figure}
\includegraphics[scale=0.6]{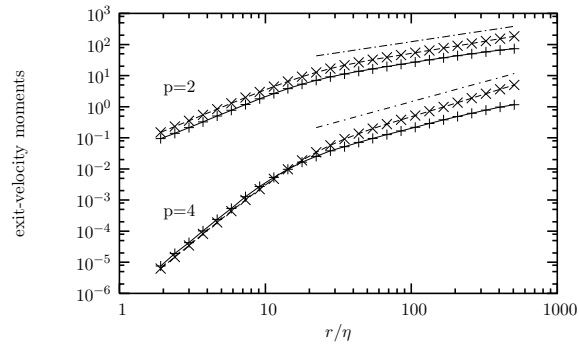}
\caption{The second and fourth order exit velocities for the
  longitudinal $u_{\|}$ ($\times$ symbol) and transverse $u_{\bot z}$
  ($+$ symbol) relative velocities for $r_0=1.2\eta$ with $\rho=1.25$.
  The dot-dashed line are the multifractal prediction for the second
  and fourth order moments.  }
\label{fig:exit-moments}
\end{figure}

We examine the intermittency of the exit velocities by considering
their second and fourth order moments. Since the exit velocity
statistics resemble Eulerian velocity statistics, we use the
multifractal formalism for Eulerian velocity structure functions to
quantify the intermittency corrections. In this way, we have a
reasonable estimate to compare with. The Eulerian structure functions
scale according to $\la \delta_r v \ra \sim (r/L_0)^{\zeta_E(p)}$ (see
e.g. \cite{Frisch}). The She-L\'ev\^eque formula for the scaling
exponents, $\zeta_E(p)$, gives $\zeta_E(2) = {0.7}$ and $\zeta_E(4) =
{1.28}$. Fig.~\ref{fig:exit-moments} shows that the second order
moment of both the longitudinal and transverse relative velocity
components scales according to the multifractal prediction. For the
fourth order moment we find that the transverse component scales well
with the multifractal prediction but the longitudinal component shows
a small discrepancy.

\section{Conclusion} \label{conclusions}
We have considered the separation process of particle pairs in
homogeneous isotropic turbulence in considerable detail. In addition
to presenting such classical but important statistics, as the pdf of
the separation distance and its second order moment (which gives us
Richardson's constant), we also considered higher moments of the
separation. Here, we found that dissipation range and integral scale
effects made a quantitative assessment of intermittency corrections in
the inertial subrange difficult. In order to get a clearer separation
between the dissipation range, inertial subrange and integral scales
we computed the statistics at fixed separations, the {\it exit time}
statistics. This provided us with an alternative method for
calculating Richardson's constant whose value agreed well with the \lq
direct' method. Moreover, these statistics allowed us to estimate
intermittency corrections quantitatively in terms of the multifractal
formalism. In agreement with Boffetta and Sokolov \cite{Boff}, we
found a small but noticeable deviation from self-similar behaviour for
those pairs that separate rapidly. Hydrodynamic turbulence is most
likely to be responsible for this rapid separation of a small
number of pairs; support for this comes from the improved scaling that
the multifractal model -- an inertial subrange model -- provides for
those particles that separate rapidly.

We also calculated the longitudinal and transverse components of the
relative velocity as a function of both travel time and the
separation, that is, the {\it exit velocities}. Analogous to the
separation statistics, we found that dissipation range and integral
scale effects made a quantitative assessment of intermittency
corrections difficult in the inertial subrange for the Lagrangian
statistics computed as a function of time. However, the fixed scale
approach showed a clear separation of scales with the velocities
scaling like the Eulerian velocity structure functions (for moments up
to order four). These statistics allowed us to quantify intermittency
corrections in terms of the multifractal model for Eulerian velocity
structure functions. However, we also noted a small but significant
difference with the true Eulerian statistics for the case of the
longitudinal exit velocity.  

\acknowledgments This work has been partially supported by the EU
under the research training networks HPRN-CT-2000-00162 "Nonideal
turbulence" and HPRN-CT-2002-00300 "Stirring and Mixing". Numerical
simulations have been performed at CINECA (INFM parallel computing
initiative and keyproject ``Lagrangian Turbulence''). We also thank
the ``Centro Ricerche e Studi Enrico Fermi'' and N.~Tantalo for
partial numerical support.


\newpage
\begin{table*}[t]
\begin{tabular}{ccccccccccccc}
$R_\lambda$ & $u_{rms}$ & $\varepsilon$ & $\nu$ & $\eta$ & $L_0$ & $T_E$
& $T_L$ & $\tau_\eta$ & $T$ & $\dd x$ & $N^3$ & $N_p$ \\ 
\hline 183 & 1.5(1) & 0.88(8) & 0.00205 &0.01 & 3.14 & 2.1 
& 1.3 & 0.048 & 5 & 0.012 & $512^3$ & 0.96$\cdot 10^{6}$  \\ 
284 & 1.7(1) & 0.81(8) & 0.00088 &0.005& 3.14 & 1.8 
& 1.2 & 0.033 & 4.4 & 0.006 & $1024^3$ & 1.92 $\cdot 10^{6}$ \\ \hline
\end{tabular}
\caption{Parameters of the numerical simulations: Taylor scale Reynolds 
number $R_\lambda$, root-mean-square velocity $u_{rms}$, mean energy
dissipation $\varepsilon$, viscosity $\nu$, Kolmogorov length scale
$\eta=(\nu^3/\varepsilon)^{1/4}$, integral scale $L_0$, large-eddy
turnover time $T_E = L_0/u_{rms}$, Lagrangian velocity autocorrelation
time $T_L$, Kolmogorov time scale $\tau_\eta=(\nu/\varepsilon)^{1/2}$,
total integration time $T$, grid spacing $\dd x$, resolution $N^3$
and the number of Lagrangian tracers $N_p$.}
\label{tab1}
\end{table*}

\newpage

\end{document}